\documentclass[12pt]{article}
\usepackage{epsfig}
\usepackage{amsmath}
\usepackage{amssymb}
\usepackage{amsfonts}

\usepackage{rotating}
\def\br(#1,#2){\left\langle#1#2\right\rangle}
\def\sq(#1,#2){\left[#1#2\right]}
\def\s(#1,#2){s_{#1 #2}}
\def\t(#1,#2,#3){s_{#1 #2 #3}}

\begin{document}

\begin{titlepage}

\hspace*{\fill}\parbox[t]{5.2cm}
{
FERMILAB-PUB-11-325-T \\
NSF-KITP-11-134\\
\today} \vskip 2cm
\begin{center}
  {\Large \bf NLO QCD predictions for $W+1$ jet and $W+2$ jet production with at least one $b$ jet at the 7 TeV LHC} \\
  \medskip \bigskip\bigskip\bigskip\bigskip {\large {\bf
      F.~Caola},$^{1,2,3}$ {\bf J.~M.~Campbell},$^1$
    {\bf F.~Febres Cordero},$^4$ {\bf L.~Reina},$^{5,7}$ and
    {\bf D.~Wackeroth}$^{6,7}$} \\
  \bigskip\bigskip\medskip
  $^{1}$Theoretical Physics Department, Fermi National Accelerator Laboratory \\
  P.~O.~Box 500, Batavia, IL\ \ 60510 \\
  \bigskip $^{2}$Dipartimento di Fisica, Universit\`a di Milano and
  INFN, Sezione di
  Milano,\\
  Via Celoria 16, I-20133 Milano, Italy \\
  \bigskip
   $^{3}$ Department of Physics and Astronomy, Johns Hopkins University, Baltimore, MD\\
  21218-2686 \\
  \bigskip
  $^{4}$Departamento de Fisica, Universidad Simon Bolivar \\ Caracas 1080A, Venezuela \\
  \bigskip
  $^{5}$Physics Department, Florida State University, Tallahassee, FL\\ 32306-4350 \\
  \bigskip 
$^{6}$Department of Physics, SUNY at Buffalo, Buffalo, NY\\
  14260-1500 \\ 
\bigskip
$^{7}$KITP, University of California Santa Barbara, CA 93106-4030, USA\\
\bigskip
\end{center}
\vfill

\begin{abstract}
  We calculate the production of a $W$ boson in association with up to
  two jets including at least one $b$-jet to next-to-leading order
  (NLO) in QCD at the CERN Large Hadron Collider with 7~TeV
  center-of-mass energy.  Both exclusive and inclusive event cross
  section and $b$-jet cross sections are presented.  The calculation
  is performed consistently in the five-flavor-number scheme where both
  $q\bar q'$ and $bq$ ($q\ne b$) initiated parton level processes are
  included at NLO QCD. We study the residual theoretical uncertainties
  of the NLO predictions due to the renormalization and factorization
  scale dependence, to the uncertainty from the parton distribution
  functions, and to the values of $\alpha_s$ and the bottom-quark mass.
\end{abstract}

\end{titlepage}

\section{Introduction}\label{sec:intro}

The study of $W$ boson production in association with one and two $b$
jets at both the Fermilab Tevatron collider ($p\bar{p}$) and the CERN
Large Hadron Collider (LHC, $pp$) has many interesting experimental
and theoretical facets. On the experimental side, these processes are
backgrounds to $WH$ production with the Higgs boson decaying to $b$
quarks, to single top and top-pair production, and to many new physics
searches.  On the theoretical side, these processes offer an
interesting testing ground for calculational techniques involving
heavy quarks with a non-negligible initial-state parton
density. Predictions for $W$ boson production in association with $b$
quarks are available at higher order in QCD using various
calculational techniques (four-flavor~\cite{Ellis:1998fv,FebresCordero:2006sj,Badger:2010mg,Frederix:2011qg}
and five-flavor number schemes~\cite{Campbell:2006cu}) and
approximations (massless~\cite{Ellis:1998fv,Campbell:2006cu} and
massive $b$
quarks~\cite{FebresCordero:2006sj,Badger:2010mg,Frederix:2011qg}).
Recently, NLO fixed-order calculations of the $q\bar q' \to W b\bar b$
parton-level process with massive $b$ quarks have been interfaced with
parton-shower Monte Carlo programs within the
POWHEG~\cite{Oleari:2011ey} and MC@NLO~\cite{Frederix:2011qg}
frameworks.

\begin{figure}[ht]
\begin{center}
\epsfxsize=5cm \epsfbox{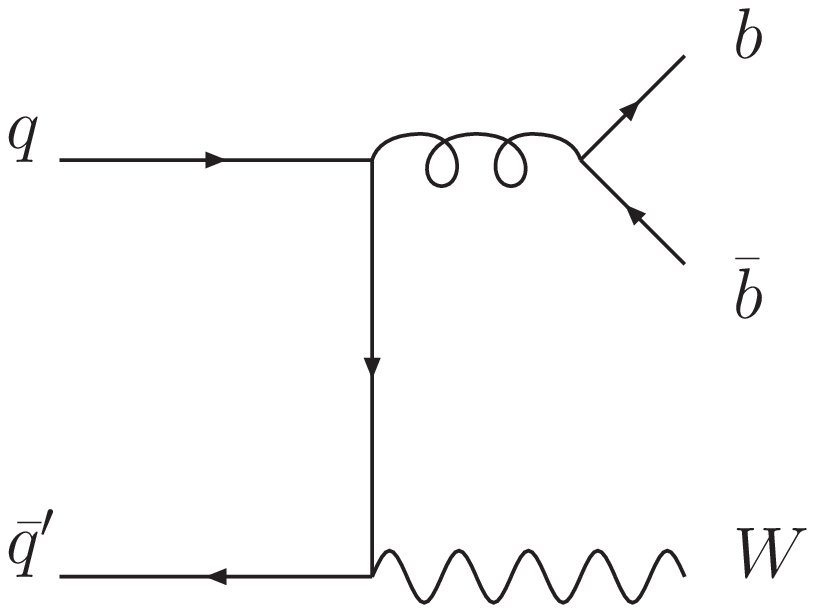} \hspace*{1cm}
\epsfxsize=5cm \epsfbox{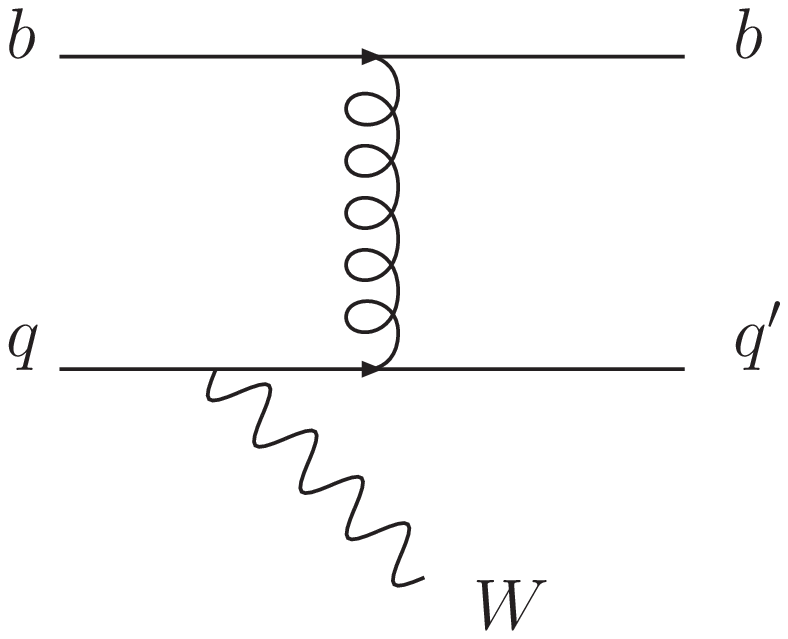} \\
(a) \hspace*{6cm} (b)
\end{center}
\caption{Leading-order parton-level processes for the production of a $W$
  boson and one or two jets with at least one $b$ jet.} \label{fig:Wbb}
\end{figure}

\begin{figure}[ht]
\begin{center}
\vspace*{.2cm} \hspace*{0cm} \epsfxsize=7cm \epsfbox{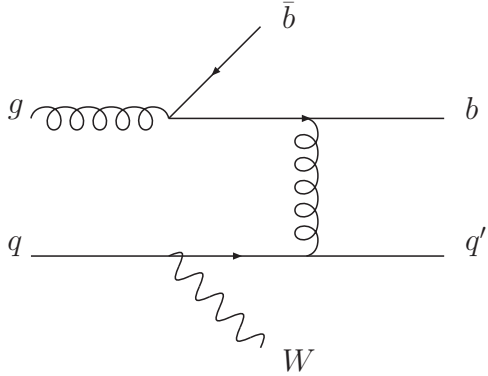}
\vspace*{-.8cm}
\end{center}
\caption{A parton-level process contributing to $Wbj$ production that appears
  at NLO in the calculation of ${\cal O}(\alpha_s)$ corrections to $q\bar
  q'\to W b \bar b$. This process is also equivalent to the LO
  $b$-quark initiated process of Fig.~\ref{fig:Wbb}(b), with the $b$ quark
  originating from collinear $g\to b \bar b$ splitting.  The consistent
  treatment of this process in the combination of the two NLO calculations is
  described in Section~\ref{sec:theory}.} \label{fig:Wbbj}
\end{figure}

In this context, the predictions for $W+1$ jet and $W+2$ jet
production with at least one $b$ jet include processes where $b$
quarks can have low transverse momentum so that finite $b$-quark mass
effects become important.  Assuming only massless quarks and gluons in
the initial state (i.e. working in a four-flavor-number scheme), this
signature can only originate from the diagram in
Fig.~\ref{fig:Wbb}(a), i.e. from $q\bar q' \to Wb\bar b$, and its
higher-order corrections.  The calculation of NLO QCD corrections to
$q\bar q' \to Wb\bar b$ of Fig.~\ref{fig:Wbb}(a) with massive $b$
quarks has been provided in
~\cite{FebresCordero:2006sj,Badger:2010mg,Frederix:2011qg}, and made
available in MCFM~\cite{MCFM}. It exhibits interesting theoretical
features.  In particular, large logarithms of the form
$\alpha_s\log(m_b/\mu)$ (where $\mu$ is a scale of the order of the
maximum $b$-quark transverse momentum) originate from the splitting of
a gluon into two almost collinear bottom quarks. This happens for the
first time in the parton-level process $qg \to Wb(\bar b) q$ (where
$(\bar b)$ denotes an untagged low $p_T$ $\bar b$ quark) depicted in
Fig.~\ref{fig:Wbbj}.  This process arises as part of the NLO QCD
corrections to $q \bar q' \to Wb\bar b$, but it is intrinsically a
tree-level process. As such it exhibits a large renormalization
($\mu_R$) and factorization ($\mu_F$) scale dependence and, because it
is enhanced by large logarithms of the form $\alpha_s \log(m_b/\mu)$,
it potentially introduces a large systematic uncertainty in the
calculation, that could be tamed only by a complete
next-to-next-to-leading (NNLO) calculation of $q\bar q' \to Wb\bar
b$. A clever way to reduce this problem is to introduce a $b$-quark
parton distribution function
(PDF)~\cite{Aivazis:1993pi,Collins:1998rz}, defined purely
perturbatively as originating from gluon splitting. In this way, the
scale evolution of the $b$-quark PDF resums the large logarithms
originating at each order and provides a more stable, although
approximate solution.  In this approach, the LO process is considered
to be $bq \to W b q'$, as shown in Fig.~\ref{fig:Wbb}(b), where the
$b$-quark PDF is generated perturbatively from the gluon PDF and the
Altarelli-Parisi splitting function for $g\to b\bar b$ splitting, and
the $\bar b$ is assumed to have too low a $p_T$ to be observable. This
results in exactly the process shown in Fig.~\ref{fig:Wbbj} with a
low-$p_T$ $b$ quark. In this approach, the $b$ quark is treated as
massless in the hard scattering process $qb\to Wbq'$, the so-called
(simplified) ACOT scheme~\cite{Aivazis:1993pi,Collins:1998rz}, and its
mass only appears as a collinear regulator in the initial $g\to b\bar b$
splitting function. The resulting logarithms $\alpha_s\ln(m_b/\mu_F)$
are resummed via DGLAP evolution of the $b$-quark PDF.  The NLO QCD
calculation of $qb\to Wbq'$ has been performed in
Ref.~\cite{Campbell:2006cu} and made available in MCFM~\cite{MCFM}.
In fact, as explained in~\cite{Campbell:2008hh}, the two tree level processes, $q\bar
q' \to Wb\bar b$ and $qb\to Wbq'$ and their $O(\alpha_s)$
corrections can be combined, as long as sufficient care is taken to subtract
logarithmic terms that would otherwise be double counted.

In this paper we will combine NLO QCD calculations of $q\bar q'\to Wb\bar b$
and $qb\to Wbq'$ parton level processes including $b$-quark mass effects to
provide precise predictions for $W+1$ jet and $W+2$ jet production with at
least one $b$ jet at the 7 TeV LHC. The choice of the experimental signature,
jet algorithm, and kinematic cuts has been made according to ATLAS
specifications~\cite{atlas}. We will closely follow Ref.~\cite{Campbell:2008hh} where a
consistent combination of these two NLO calculations has been performed for
the first time to provide predictions for the production of a $W$ boson and
one $b$-jet.
It is interesting to note that the calculation of Ref~\cite{Campbell:2008hh}
has been compared with a measurement of the $b$-jet cross section of $W$ boson
production in association with one and two $b$ jets by the CDF collaboration at
the Tevatron~\cite{Aaltonen:2009qi}. This comparison found a discrepancy of
about two standard deviations~\cite{Cordero:2010qn,jfl}.

After a brief presentation of the theoretical framework in
Section~\ref{sec:theory}, we will discuss NLO QCD predictions and their
residual uncertainties for the 7~TeV LHC in Section~\ref{sec:results} and
present our conclusions in Section~\ref{sec:conclusions}.

\section{Theoretical Framework}\label{sec:theory}

The predictions presented in this paper are based on the combination
of NLO QCD calculations of the $q\bar q' \to W b \bar
b$~\cite{FebresCordero:2006sj,Badger:2010mg,MCFM} and $bq \to W
bq$~\cite{Campbell:2006cu} parton-level processes, as presented in
Ref.~\cite{Campbell:2008hh} and implemented in MCFM~\cite{MCFM} (where
the leptonic $W$ decay is included), and we refer
to~\cite{Campbell:2008hh} for more details.

In the NLO QCD calculation of the $q\bar q' \to W b \bar b$ process
the $b$ quark is considered to be massive, and only light quarks
($q\ne b$) are considered in the initial state, i.e. the so-called
four-flavor number scheme (4FNS) is used. In the NLO QCD calculation
of the $bq \to W b q'$ process the $b$-quark mass is only kept as
regulator of the collinear singularity while it is neglected in the
hard process so that the hadronic cross section is obtained as
follows,
\begin{equation}
\sigma_{bq}^{NLO}=\int dx_1 dx_2 b(x_1,\mu) \left[
\sum_q q(x_2,\mu_F) \hat\sigma_{bq}^{NLO}(m_b=0)+
g(x_2,\mu_F) \hat \sigma_{bg}^{LO}(m_b=0) \right]\,\,\,.
\end{equation}
An approximate solution of the DGLAP evolution equation for the
$b$-quark PDF $b(x,\mu_F)$ with initial condition $b(x,\mu_F)=0$ at
$\mu_F=m_b$ exhibits the collinear logarithm at leading order in
$\alpha_s$ as follows~\cite{Aivazis:1993pi,Collins:1998rz},
\begin{equation}
\tilde b(x,\mu_F)=\frac{\alpha_s(\mu_R)}{\pi} \log\left(\frac{\mu_F}{m_b}\right) \int_x^1
\frac{dz}{z} P_{qg}(z) g\left(\frac{x}{z},\mu_F\right)\,\,\,.
\end{equation}
When combining the NLO calculation of this process with the NLO
calculation of $q\bar q'\to Wb\bar b$ this contribution has to be
subtracted in order to avoid double counting of the process of
Fig.~\ref{fig:Wbbj} which is already included in the 4FNS NLO QCD
calculation. The full five-flavor number scheme (5FNS) result at NLO
QCD, including an all order resummation of collinear initial-state
logarithms via DGLAP evolution, is then
obtained schematically as follows,
\begin{eqnarray}
\label{eq:5fns}
\sigma_{\mathrm{Full}}^{NLO}&=& \sigma_{\mathrm4FNS}^{NLO}(m_b \ne 0) +  \sigma_{bq}^{NLO} \nonumber\\
&-& \sum_q\int dx_1 dx_2 \tilde b(x_1,\mu_F) q(x_2,\mu_F) \hat \sigma_{bq}^{LO}(m_b=0)\,\,\,.
\end{eqnarray}
In fact, the situation is slightly more complicated because the NLO computations of the $q \bar q' \to W b \bar b$ and of 
the $b q \to W b q'$ processes are made in two different schemes, one in the $\overline{\textrm{MS}}$ scheme and the other
in a decoupling scheme. Hence, in Eq.~(\ref{eq:5fns}) a scheme change is also assumed, for which
we refer the reader to the literature~\cite{Aivazis:1993pi,Collins:1998rz,Cacciari:1998it} for further details.
This said, we now present the sub-processes relevant for our analysis. 
In detail, $\sigma_{4FNS}^{NLO}$ and $\sigma_{bq}^{NLO}$ in this paper include the following parton level processes:
\vskip 0.5cm 
\begin{tabular}{c@{\hspace{1cm}}l}
& $q\bar q'\to Wb\bar b$ at tree level [Fig.~\ref{fig:Wbb}(a)] and one loop ($m_b\neq 0$) \\
$\sigma_{4FNS}^{NLO}$ & $q\bar q'\to Wb\bar bg$ at tree level ($m_b\neq 0$) \\
& $gq\to Wb\bar bq'$ at tree level (Fig.~\ref{fig:Wbbj}), ($m_b\neq 0$) \\ 
& \\
& $bq\to Wbq'$ at tree level [Fig.~\ref{fig:Wbb}(b)] and one loop ($m_b=0$) \\
$\sigma_{bq}^{NLO}$ & $bq\to Wbq'g$ at tree level ($m_b=0$) \\
& $bg\to Wbq'\bar q$ at tree level ($m_b=0$) \\
\end{tabular}
\vskip 0.5cm
We present results for $\sigma_{\mathrm{Full}}^{NLO}$ and
$\sigma_{4FNS}^{NLO}$ separately in
Tables~\ref{tab:inclusive}-\ref{tab:exclusive} for the signatures
described in the following section.
We leave a full discussion of these results until Section~\ref{sec:results}, but here simply note
that the scale dependence of the
difference $\sigma_{\mathrm{Full}}^{NLO}-\sigma_{4FNS}^{NLO}$ clearly
shows the impact of the initial-state collinear logarithms. The difference is
negligible for scales of the order of the $b$-quark mass but can
amount to about 40\% of $\sigma_{\mathrm{Full}}^{NLO}$ for $\mu \approx
360$~GeV.

Note that we do not include contributions to $Wbj$ production which arise from
$c\to Wb$ transitions (we assume $V_{cb}=0$), since these contributions are
suppressed by the smallness of $V_{cb}$ and the charm quark PDF. For instance,
the dominant contribution to $Wbj$ production at the LHC when considering
$|V_{cb}|=0.04$ is expected to be $cg \to W bg$. Using the setup of
Table~\ref{tab:parameters} we found for the inclusive $W^+b$ event cross
section at LO QCD with $\mu=\mu_0$: $290 \times |V_{cb}|^2=0.5$ pb, which is
about 1\% of the result presented in Table~\ref{tab:inclusive}.

In the calculation of $\sigma_{\mathrm{Full}}^{NLO}$ in the full 5FNS of
Eq.~(\ref{eq:5fns}) we assume the number of light flavors to be $n_{lf}=5$ in
both the running of $\alpha_s(\mu_R)$ and in the determination of the
one-loop gluon self energy, $\Sigma_{gg}$, i.e. we only decouple
the top quark from the running of $\alpha_s$ in the modified $\overline{\rm
  MS}$ scheme. This choice is motivated by the fact that we usually choose 
renormalization scales $\mu_R$ considerably larger than $m_b$.
Alternatively, one can choose to include the $b$-quark mass in the calculation
of $\Sigma_{gg}$ (as done in the implementation of $q\bar q'\to Wb\bar b$ in
POWHEG~\cite{Oleari:2011ey}) and/or also decouple the $b$ quark ($n_{lf}=4$). The different
treatments generally result in differences of about a few per cent in the cross sections
presented in Section~\ref{sec:results}.

\section{Results}\label{sec:results}

Predictions are provided for $W+1$ jet and $W+2$ jet production where at least
one jet is a $b-$jet as will be measured by ATLAS at the 7~TeV
LHC~\cite{atlas}.  Jets are clusters of partons built using the anti-$k_T$
algorithm which passed the kinematic cuts specified in
Table~\ref{tab:parameters}. In the following, $b$ denotes a jet containing one $b$ quark or one
$\bar{b}$ antiquark, while $(bb)$ denotes a jet containing a $b$ and $\bar b$
quark. $b$ and $(bb)$ jets may also contain a light parton.  $j$ labels a jet
without $b$ quarks. We will provide predictions for event and $b$-jet cross sections for the following signatures:
\begin{itemize}
\item 
$Wb$ inclusive: one and two-jet events with $b$ jets containing a single $b$, i.e. $Wb+Wbj+Wb\bar b$. 
\item 
$W(bb)$ inclusive: one and two-jet events with one $(bb)$ jet,
i.e. $W(bb)+W(bb)j$.
This signature can only result from processes contributing to $
\sigma_{4FNS}^{NLO}$ listed in Section~\ref{sec:theory}. 
\item
$Wb$ exclusive: one-jet events with one $b$ jet containing a single $b$, i.e. $Wb$.
\item 
$W(bb)$ exclusive: one-jet events with one $(bb)$ jet, i.e. $W(bb)$ 
This signature can only result from the processes contributing to $
\sigma_{4FNS}^{NLO}$ listed in Section~\ref{sec:theory}. 
\end{itemize}
The event and $b$-jet cross sections have been obtained consistently at NLO in the 5FNS
following Ref.~\cite{Campbell:2008hh} and as briefly described in
Section~\ref{sec:theory}.  

If not stated otherwise, all results are obtained assuming $\mu_R=\mu_F=\mu$,
where $\mu_R$ and $\mu_F$ denote the renormalization and factorization scales
respectively. We vary $\mu$ between $\mu_0/4$ and $4\mu_0$ with $\mu_0=M_W+2
m_b$. Results for the event cross sections corresponding to the four
signatures described above are given in
Tables~\ref{tab:inclusive}-\ref{tab:exclusive}, where we consider non-decaying
$W$ bosons. The theoretical uncertainty due to the scale dependence can be
estimated using these results.  Inclusive and exclusive event cross sections
for $pp\to W^\pm bX\to e^\pm\nu bX$ assuming $\mu=\mu_0$ are provided in
Table~\ref{tab:eventcut}. The results have been obtained by multiplying the
total cross sections of Tables~\ref{tab:inclusive}-\ref{tab:exclusive} with
the branching ratio $\text{BR}(W^\pm \to e^\pm \nu)=0.10805$ (labeled as ``no
cuts'') and by requiring ATLAS inspired lepton cuts, $p_T^e > 20$~GeV,
$|\eta^e|<2.5$, $p_T^\nu > 25$~GeV, $m_T^W > 40$~GeV, $R(l,j)>0.5$ (labeled as
``ATLAS cuts'').  All these cuts are implemented in the full NLO computation
including $W$ decay of MCFM~\cite{MCFM}.

The PDF uncertainties are estimated using the
NNPDF2.1~\cite{Ball:2010de}, CTEQ6.6~\cite{Nadolsky:2008zw}, and MSTW08~\cite{Martin:2009iq}
sets of PDFs as presented in Tables~\ref{tab:incerror}-\ref{tab:excerror}. 
Also shown are predictions for the event cross sections for different values of $\alpha_s$ 
obtained with NNPDF2.1~\cite{Ball:2010de}. The dependence of our predictions on the value of the
$b$-quark mass is  at the level of a few percent, as can be seen from Tables~\ref{tab:incmb}-\ref{tab:excmb}.

\begin{table} \begin{center} \caption[fake]{Kinematic cuts, jet finding
      algorithm, PDF sets and input parameters  
used in this study, if not stated otherwise. 
The kinematic cuts used to simulate the acceptance and
resolution of the detectors are chosen according to ATLAS
specifications~\cite{atlas}. }\label{tab:parameters}
\bigskip \begin{tabular} {ll}
7 TeV LHC:  $\;\;\;\;\;\,$ $p_{Tj}>25$ GeV & $|y_j|<2.1$ \\
anti$-k_T$ jet algorithm &  $p=-1, R=0.4$ \\ \\
$M_W=80.41$ GeV & $m_b=4.7$ GeV \\
LO: CTEQ6L1 & NLO: CTEQ6.6~\cite{Nadolsky:2008zw}\\
$\alpha_S^{LO}(M_Z)=0.130$ & $\alpha_S^{NLO}(M_Z)=0.118$  \\
$g_w^2=8M_W^2G_F/\sqrt 2=0.4266177$ & $G_F=1.16639\times 10^{-5}$ GeV$^{-2}$ \\
$V_{ud}=V_{cs}=0.974$ & $V_{us}=V_{cd}=0.227$ \; \;  \; ($V_{ub}=V_{cb}=0$) \\
\end{tabular} \end{center}
\end{table}

The predictions for the event cross sections for $W+1$ jet and $W+2$
jets with at least one $b$ (or $(bb)$) jet, denoted as
$\sigma_{1j+2j}$, $W+1 b$ jet (or $(bb)$ jet), denoted as
$\sigma_{1j}$, and $W+2$ jets with at least one $b$ (or $(bb)$) jet,
denoted as $\sigma_{2j}$, are provided separately in
Table~\ref{tab:wobs}. They are obtained from the results of
Tables~\ref{tab:inclusive}-\ref{tab:exclusive} as follows:
\begin{eqnarray*}
\sigma_{1j+2j}&=&
\left[\sigma_{\mathrm{event}}(Wb\,\mathrm{incl.})+\sigma_{\mathrm{event}}(W(bb)\,\mathrm{incl.}) \right] \\
\sigma_{1j}&=&
\left[\sigma_{\mathrm{event}}(Wb\,\mathrm{excl.})+\sigma_{\mathrm{event}}(W(bb)\,\mathrm{excl.}) \right] \\
\sigma_{2j}&=& \sigma_{1j+2j}-\sigma_{1j}
\end{eqnarray*}

The $b$-jet cross sections for $W+1$ jet and $W+2$ jets with at least one $b$
jet are provided separately in Table~\ref{tab:bjet}. They can be obtained from
the $Wb$ and $W(bb)$ inclusive event cross sections of
Table~\ref{tab:inclusive} when the $Wb\bar{b}$ contribution (normally included
in the $Wb$ inclusive signatures) is counted twice (since it contains two $b$
jets). More explicitly, using the $Wb\bar b$ cross section separately provided
in Table~\ref{tab:wobs} in parentheses, the $b$-jet cross section can be
obtained from the event cross sections in Table~\ref{tab:inclusive} as follows
\begin{eqnarray*}
\sigma_{b-\mathrm{jet}}&=&
\left[\sigma_{\mathrm{event}}(Wb\,\mathrm{incl.})-\sigma_{\mathrm{event}}(Wb\bar{b})\right]+
2\,\sigma_{\mathrm{event}}(Wb\bar{b})+\sigma_{\mathrm{event}}(W(bb)\,\mathrm{incl.})\\
&=&\sigma_{\mathrm{event}}(Wb\,\mathrm{incl.})+
\sigma_{\mathrm{event}}(Wb\bar{b})+\sigma_{\mathrm{event}}(W(bb)\,\mathrm{incl.})\\
&=&\sigma_{1j+2j}+\sigma_{\mathrm{event}}(Wb\bar{b})
\end{eqnarray*}
\begin{table}[t]
\begin{center}
  \caption{\label{tab:inclusive}Inclusive event cross sections (in pb),
    LHC ($\sqrt{s}=7$ TeV). No branching ratios or tagging
    efficiencies are included. The Monte Carlo integration error is 0.5\%.
  }
\bigskip\begin{tabular}{|c|c|c|c|c|c|c|c|c|} \hline\hline
& \multicolumn{2}{|c|} {$W^+ b$ incl.} & 
$W^+ (b b)$ incl. & 
\multicolumn{2}{|c|} {$W^- b$ incl.} & 
$W^- (b b)$ incl. \\
\hline
& Full & 4FNS &  4FNS & Full & 4FNS &  4FNS\\
\hline
$\mu=\mu_0/4$ & 66.3 & 67.3& 18.6 & 40.8 &41.2 & 11.4 \\
$\mu=\mu_0/2$ & 60.4 &52.5& 13.8 & 37.2 &32.2& 8.6 \\
$\mu=\mu_0$   & 56.7 &42.6& 10.9 & 34.8 &26.3 & 6.8 \\
$\mu=2 \mu_0$ & 53.2 &35.5& 8.8 & 32.7 &21.9& 5.4 \\
$\mu=4 \mu_0$ & 50.0 &30.1& 7.4 & 30.7 &18.7 & 4.5 \\
\hline
\end{tabular}
\end{center}
\end{table}
%
%
\begin{table}[t]
\begin{center}
  \caption{\label{tab:exclusive}Exclusive event cross sections (in pb),
    LHC ($\sqrt{s}=7$ TeV). No branching ratios or tagging
    efficiencies are included. The Monte Carlo integration error is within
    0.5\%. }
\bigskip\begin{tabular}{|c|c|c|c|c|c|c|c|c|} \hline\hline
&\multicolumn{2}{|c|} {$W^+ b$ excl.} & 
$W^+ (b b)$ excl. & 
\multicolumn{2}{|c|} {$W^- b$ excl.} & 
$W^- (b b)$ excl. \\
\hline
& Full & 4FNS &  4FNS & Full & 4FNS &  4FNS\\
\hline
$\mu=\mu_0/4$ & 36.7 &36.9 & 9.4 & 22.8 &22.3& 5.7   \\
$\mu=\mu_0/2$ & 35.3 &35.2& 7.8 & 21.8 &21.5& 4.9  \\
$\mu=\mu_0$   & 33.9 &26.2& 6.7 & 20.7 &16.2& 4.3 \\
$\mu=2 \mu_0$ & 32.2 &22.8& 5.9 & 19.8 &14.2& 3.7 \\
$\mu=4 \mu_0$ & 30.3 &19.9& 5.2 & 18.8 &12.5& 3.3 \\
\hline
\end{tabular}
\end{center}
\end{table}

\begin{table}
\centering
\begin{tabular}{|c|c|c|c|c|c|c|c|c|}
\hline
\hline
& 
\multicolumn{2}{|c|} {$W b$ incl.} & 
$W (b b)$ incl. & 
\multicolumn{2}{|c|} {$W b$ excl.} & 
$W (b b)$ excl. \\
\hline
& 4FNS & Full &  4FNS & 4FNS & Full &  4FNS\\
\hline
$W^+$ no cuts    & 4.6& 6.1& 1.2& 2.8& 3.7& 0.7\\
$W^+$ ATLAS cuts  & 2.2 & 2.8 
                 & 0.5 
                 & 1.3 & 1.7
                 & 0.3\\ \hline
$W^-$ no cuts    &2.8 &3.8 &0.7 &1.8 & 2.2& 0.5\\
$W^-$ ATLAS cuts & 1.3 & 1.6
                 & 0.3
                 & 0.8 & 1.0
                 & 0.2\\
\hline
\end{tabular}
  \caption{\label{tab:eventcut}Inclusive and exclusive event cross sections (in pb),
    LHC ($\sqrt{s}=7$ TeV), for $pp\to W^\pm bX\to e^\pm\nu bX$ (with
    $\mu=\mu_0$ and CTEQ6.6~\cite{Nadolsky:2008zw}). The ``no cuts''
    result is obtained by multiplying the total cross sections of
    Table~\ref{tab:inclusive} and Table~\ref{tab:exclusive}, respectively, with
    the branching ratio $\text{BR}(W^\pm \to e^\pm \nu)=0.10805$. The ``ATLAS cuts''
    result is obtained by requiring $p_T^e > 20$~GeV, $|\eta^e|<2.5$, 
    $p_T^\nu > 25$~GeV, $m_T^W > 40$~GeV, $R(l,j)>0.5$. 
    All these cuts are implemented in the full NLO computation including $W$ decay
    of MCFM~\cite{MCFM}.}
\end{table}

%
%

\begin{table}
\centering
\begin{tabular}{|c|c|c|c|c|c|c|c|c|}
\hline
\hline
& 
\multicolumn{2}{|c|} {$W^+ b$ incl.} & 
$W^+ (b b)$ incl. & 
\multicolumn{2}{|c|} {$W^- b$ incl.} & 
$W^- (b b)$ incl. \\
\hline
& 4FNS & Full &  4FNS & 4FNS & Full &  4FNS\\
\hline
NNPDF2.1~\cite{Ball:2010de} & 44.1 & 59.2 $\pm$ 0.7
         & 11.4 $\pm$ 0.1  
         & 27.6 & 36.2 $\pm$ 0.6
         & 7.1 $\pm$ 0.1\\
CT10~\cite{Lai:2010vv}     & 42.1 & 56.1 $\pm$ 1.5
         & 10.9 $\pm$ 0.3
         & 26.5 & 34.7 $\pm$ 1.0
         & 6.9 $\pm$ 0.2\\
CTEQ6.6~\cite{ Nadolsky:2008zw} & 42.6 & 56.8 $\pm$ 1.4
            & 10.9 $\pm$ 0.2
            & 26.3 & 34.6 $\pm$ 1.0  
            & 6.8 $\pm$ 0.2\\
MSTW2008~\cite{Martin:2009iq} & 44.2 & 59.8 $\pm$ 0.7
         & 11.5 $\pm$  0.1
         & 28.6 & 37.9 $\pm$ 0.5
         & 7.4 $\pm$ 0.1\\
\hline
$\alpha_s(M_Z)=0.114$ & 39.2 & 52.0
         & 10.1 
         & 24.3 & 31.6
         & 6.3\\
$\alpha_s(M_Z)=0.124$ & 49.6 & 66.9
         & 12.8
         & 31.0 & 41.3
         & 8.0\\
\hline
\end{tabular}
\caption{Inclusive event cross sections (in pb) for different PDF sets
  including PDF uncertainties (NNPDF2.1: full error set. 
CT10~\cite{Lai:2010vv}/MSTW2008~\cite{Martin:2009iq}: central set), and for different values of $\alpha_s$ 
obtained with NNPDF2.1~\cite{Ball:2010de} (with $\mu_R=\mu_F=\mu_0$). }\label{tab:incerror} 
\end{table}

\begin{table}
\centering
\begin{tabular}{|c|c|c|c|c|c|c|c|c|}
\hline
\hline
& 
\multicolumn{2}{|c|} {$W^+ b$ excl.} & 
$W^+ (b b)$ excl. & 
\multicolumn{2}{|c|} {$W^- b$ excl.} & 
$W^- (b b)$ excl. \\
\hline
& 4FNS & Full &  4FNS & 4FNS & Full &  4FNS\\
\hline
NNPDF2.1~\cite{Ball:2010de}   & 26.5 & 34.9 $\pm$ 0.5 
           & 7.0 $\pm$ 0.1
           & 16.9 & 21.6 $\pm$ 0.4
           & 4.4 $\pm$ 0.1\\
CT10~\cite{Lai:2010vv} & 26.1 & 33.6 $\pm$ 1.0
            & 6.8 $\pm$ 0.2
            & 16.3 & 20.8 $\pm$ 0.7
            & 4.2 $\pm$ 0.2\\
CTEQ6.6~\cite{ Nadolsky:2008zw} & 26.2 & 33.9 $\pm$ 0.9 
            & 6.7 $\pm$ 0.2
            & 16.2 & 20.8 $\pm$ 0.7 
            & 4.3 $\pm$ 0.2\\
MSTW2008~\cite{Martin:2009iq}  & 27.4 & 35.6 $\pm$ 0.5
                 & 7.1 $\pm$ 0.1
                 & 17.6 & 22.4 $\pm$ 0.3
                 & 4.6 $\pm$ 0.1\\
\hline
$\alpha_s(M_Z)=0.114$ & 24.1 & 31.1
            & 6.2
            & 14.9 & 18.9
            & 3.8\\
$\alpha_s(M_Z)=0.124$ & 30.4 & 39.7
            & 7.8
            & 19.0 & 24.5
            & 4.9\\
\hline
\end{tabular}
\caption{Exclusive event cross sections (in pb) for different PDF sets
  including PDF uncertainties (NNPDF2.1: full error set. 
CT10~\cite{Lai:2010vv}/MSTW2008~\cite{Martin:2009iq}: central set), and for different values of $\alpha_s$ 
obtained with NNPDF2.1~\cite{Ball:2010de} (with $\mu_R=\mu_F=\mu_0$).}  \label{tab:excerror} 
\end{table}

\begin{table}
\centering
\begin{tabular}{|c|c|c|c|c|c|c|c|c|}
\hline
\hline
& 
\multicolumn{2}{|c|} {$W^+ b$ incl.} & 
$W^+ (b b)$ incl. & 
\multicolumn{2}{|c|} {$W^- b$ incl.} & 
$W^- (b b)$ incl. \\
\hline
& 4FNS & Full &  4FNS & 4FNS & Full &  4FNS\\
\hline
$m_b=4.2~$GeV & 44.8 & 58.4
              & 12.8
              & 27.7 & 35.8
              & 7.9\\
$m_b=5.0$~GeV & 40.8 & 55.7
              & 10.0
              & 25.2 & 34.0
              & 6.2\\
\hline
\end{tabular}
\caption{Inclusive event cross sections (in pb)  for different values of the
  $b$ quark  mass ($m_b$) obtained with the CTEQ6.6 PDF set~\cite{Nadolsky:2008zw} and $\mu_R=\mu_F=\mu_0$.}\label{tab:incmb}
\end{table}

\begin{table}
\centering
\begin{tabular}{|c|c|c|c|c|c|c|c|c|}
\hline
\hline
& 
\multicolumn{2}{|c|} {$W^+ b$ excl.} & 
$W^+ (b b)$ excl. & 
\multicolumn{2}{|c|} {$W^- b$ excl.} & 
$W^- (b b)$ excl. \\
\hline
& 4FNS & Full &  4FNS & 4FNS & Full &  4FNS\\
\hline
$m_b=4.2~$GeV & 27.8 & 35.3
              & 8.0
              & 17.2 & 21.5
              & 5.0\\
$m_b=5.0$~GeV & 25.0 & 33.3
              & 6.1
              & 15.5 & 20.2
              & 3.8\\
\hline
\end{tabular}
\caption{Exclusive event cross sections (in pb) for different values of the
  $b$ quark mass ($m_b$) obtained with the CTEQ6.6 
  PDF set~\cite{Nadolsky:2008zw} and  $\mu_R=\mu_F=\mu_0$.}\label{tab:excmb}
\end{table}


\begin{table}[t]
\begin{center}
  \caption{Event cross sections (in pb), LHC ($\sqrt{s}=7$ TeV) for
    $W+1$ and $W+2$ jet production with at least one $b$ jet, $W+1$
    $b$ jet, and $W+2$ jets with at least one $b$ jet (where here $b$
    jet denotes a jet with a single $b$ or $(bb)$ pair). The event
    cross sections for $Wb\bar b$ are provided separately in
    parentheses. No branching ratios or tagging efficiencies are
    included. The Monte Carlo integration error is within
    0.5\%. }\label{tab:wobs}
\bigskip\begin{tabular}{|c|c|c|c|c|c|c|} \hline\hline
 & $W^+_{1j+2j}$ & $W^+_{1j}$ & $W^+_{2j}$  & $W^-_{1j+2j}$ & $W^-_{1j}$ & $W^-_{2j}$ \\
\hline
$\mu=\mu_0/4$ & 84.9 [5.6] & 46.1 & 38.8   & 52.2 [3.2] & 28.5  & 23.7   \\
$\mu=\mu_0/2$ & 74.2 [5.3] & 43.1  & 31.1   & 45.8 [3.1] & 26.7 & 19.1   \\
$\mu=\mu_0$   & 67.6 [5.0]& 40.6 & 27.0  & 41.6 [2.9]& 25.0 & 16.6  \\
$\mu=2 \mu_0$ & 62.0 [4.6]&  38.1& 23.9 & 38.1 [2.7]& 23.5 &  14.6 \\
$\mu=4 \mu_0$ & 57.4 [4.2]&  35.5 & 21.9   & 35.2 [2.5]& 22.1 &  13.1 \\
\hline
\end{tabular}
\end{center}
\end{table}
\begin{table}[t]
\begin{center}
  \caption{$b$-jet cross sections
    (in pb) for $W+1$ and $W+2$ jet production where at least one jet is a $b$ jet, LHC ($\sqrt{s}=7$ TeV). No
    branching ratios or tagging efficiencies are included. The Monte Carlo
    integration error is 0.5\%. }
\label{tab:bjet}
\bigskip\begin{tabular}{|c|c|c|} \hline\hline
& $\sigma_{b-\mathrm{jet}}(W^+)$ & $\sigma_{b-\mathrm{jet}}(W^-)$ \\
\hline
$\mu=\mu_0/4$ & 90.5 & 55.4   \\
$\mu=\mu_0/2$ & 79.5 & 48.9   \\
$\mu=\mu_0$   & 72.6 &  44.5  \\
$\mu=2 \mu_0$ & 66.6 &  40.8\\
$\mu=4 \mu_0$ & 61.6 & 37.7  \\
\hline
\end{tabular}
\end{center}
\end{table}

\section{Conclusions}\label{sec:conclusions}

In this paper we have computed the cross section for
the production of a $W$ boson in association with up to
two jets, including at least one $b$-jet, at the $7$~TeV LHC.
The calculation consistently combines next-to-leading order corrections
to the parton level processes
$q\bar q' \to W b \bar b$~\cite{FebresCordero:2006sj,Badger:2010mg,MCFM}
and $bq \to W bq$~\cite{Campbell:2006cu} according to the
procedure presented in Ref.~\cite{Campbell:2008hh}.
We have particularly focussed on performing a systematic study of our
prediction, considering a number of sources of theoretical uncertainty,
under a set of cuts that will be used by the ATLAS
collaboration~\cite{atlas}.
  
Our results can be summarized as follows, where we have put together
all sources of uncertainty considered in this paper.
The event cross section for $W$ production with one or two jets with at
least one $b$ jet at NLO QCD at the LHC (7 TeV) as will be measured by ATLAS~\cite{atlas} is,
\[\sigma_{1j+2j}(W^++W^-)=109.2 ~{}^{+27.9}_{-16.6} \, ({\rm scale}) ~{}^{+7.4}_{-1.9} \, ({\rm PDF})
 ~{}^{+5.7}_{-3.3} \, (m_b) ~{\rm pb}\]
The central value corresponds to the CTEQ6.6 PDF set, $\mu = \mu_0$ and $m_b=4.7$~GeV.
In the assessment of the theoretical uncertainty we have considered a very conservative
scale variation from $\mu_0/4$ to $4\mu_0$, i.e. ranging from approximately $20$ to $360$~GeV.
The PDF uncertainty is assessed by conmparing the nominal CTEQ6.6 prediction with the results
obtained for CT10 and MSTW08, while $m_b$ is varied from $4.2$ to $5$~GeV.


\section*{Acknowledgments}

We would like to thank Tobias Golling and Andrea Messina from the ATLAS
collaboration for many fruitful discussions and for providing us all the
information necessary to obtain the results presented in this paper.  L.R. and
D.W. thank the Kavli Institute for Theoretical Physics (KITP) for the kind
hospitality extended to us while this work was being completed.  The work of
L.R. is supported in part by the U.S. Department of Energy under grant
DE-FG02-97IR41022.  The work of D.W. is supported in
part by the National Science Foundation under grants NSF-PHY-0547564 and
NSF-PHY-0757691.
Fermilab is operated by Fermi Research Alliance, LLC under
Contract No. DE-AC02-07CH11359 with the United States Department of Energy. 
This research was supported in part by the National Science
Foundation under Grant No. NSF PHY05-51164 and by a US-Italy Fulbright Visiting Student Researcher Fellowship.

\end{document}